\documentstyle[12pt,twoside]{article}

\def\a{\alpha}
\def\b{\beta}
\def\c{\chi}
\def\d{\delta}
\def\e{\epsilon}                
\def\f{\phi}                    
\def\g{\gamma}
\def\h{\eta}

\def\j{\psi}
\def\k{\kappa}
\def\l{\lambda}
\def\m{\mu}
\def\n{\nu}
\def\o{\omega}
\def\r{\rho}                    
\def\s{\sigma}                  
\def\t{\tau}

\def\x{\xi}

\def\D{\Delta}

\def\Q{\Theta}

\def\cb{{\cal B}}

\def\cl{{\cal L}}

\catcode`@=11

\def\un#1{\relax\ifmmode\@@underline#1\else $\@@underline{\hbox{#1}}$\relax\fi}

{\catcode`\'=\active \gdef'{{}^\bgroup\prim@s}}

\def\magstep#1{\ifcase#1 \@m\or 1200\or 1440\or 1728\or 2074\or 2488\or
        2986\fi\relax}

\font\twfvmi=cmmi10\@magscale5
    \skewchar\twfvmi='177
\font\twfvsy=cmsy10\@magscale5
    \skewchar\twfvsy='60
\font\twfvly=lasy10\@magscale5
\font\thtyrm=cmr10\@magscale6

\def\vpt{\textfont\z@\fivrm
  \scriptfont\z@\fivrm \scriptscriptfont\z@\fivrm
\textfont\@ne\fivmi \scriptfont\@ne\fivmi \scriptscriptfont\@ne\fivmi
\textfont\tw@\fivsy \scriptfont\tw@\fivsy \scriptscriptfont\tw@\fivsy
\textfont\thr@@\tenex \scriptfont\thr@@\tenex \scriptscriptfont\thr@@\tenex
\def\prm{\fam\z@\fivrm}%
\def\unboldmath{\everymath{}\everydisplay{}\@nomath
  \unboldmath\fam\@ne\@boldfalse}\@boldfalse
\def\boldmath{\@subfont\boldmath\unboldmath}%
\def\pit{\@getfont\pit\itfam\@vpt{cmti5}}%
\def\psl{\@subfont\sl\it}%
\def\pbf{\@getfont\pbf\bffam\@vpt{cmbx5}}%
\def\ptt{\@subfont\tt\rm}%
\def\psf{\@subfont\sf\rm}%
\def\psc{\@subfont\sc\rm}%
\def\ly{\fam\lyfam\fivly}\textfont\lyfam\fivly
    \scriptfont\lyfam\fivly \scriptscriptfont\lyfam\fivly
\@setstrut\rm}

\def\@vpt{}

\def\vipt{\textfont\z@\sixrm
  \scriptfont\z@\sixrm \scriptscriptfont\z@\sixrm
\textfont\@ne\sixmi \scriptfont\@ne\sixmi \scriptscriptfont\@ne\sixmi
\textfont\tw@\sixsy \scriptfont\tw@\sixsy \scriptscriptfont\tw@\sixsy
\textfont\thr@@\tenex \scriptfont\thr@@\tenex \scriptscriptfont\thr@@\tenex
\def\prm{\fam\z@\sixrm}%
\def\unboldmath{\everymath{}\everydisplay{}\@nomath
  \unboldmath\@boldfalse}\@boldfalse
\def\boldmath{\@subfont\boldmath\unboldmath}%
\def\pit{\@subfont\it\rm}%
\def\psl{\@subfont\sl\rm}%
\def\pbf{\@getfont\pbf\bffam\@vipt{cmbx6}}%
\def\ptt{\@subfont\tt\rm}%
\def\psf{\@subfont\sf\rm}%
\def\psc{\@subfont\sc\rm}%
\def\ly{\fam\lyfam\sixly}\textfont\lyfam\sixly
    \scriptfont\lyfam\sixly \scriptscriptfont\lyfam\sixly
\@setstrut\rm}

\def\@vipt{}

\def\xxxpt{\textfont\z@\thtyrm
  \scriptfont\z@\twfvrm \scriptscriptfont\z@\twtyrm
\textfont\@ne\twfvmi \scriptfont\@ne\twfvmi \scriptscriptfont\@ne\twtymi
\textfont\tw@\twfvsy \scriptfont\tw@\twfvsy \scriptscriptfont\tw@\twtysy
\textfont\thr@@\tenex \scriptfont\thr@@\tenex \scriptscriptfont\thr@@\tenex
\def\unboldmath{\everymath{}\everydisplay{}\@nomath\unboldmath
        \textfont\@ne\twfvmi \textfont\tw@\twfvsy \textfont\lyfam\twfvly
        \@boldfalse}\@boldfalse
\def\boldmath{\@subfont\boldmath\unboldmath}%
\def\prm{\fam\z@\thtyrm}%
\def\pit{\@subfont\it\rm}%
\def\psl{\@subfont\sl\rm}%
\def\pbf{\@getfont\pbf\bffam\@xxxpt{cmbx10\@magscale6}}%
\def\ptt{\@subfont\tt\rm}%
\def\psf{\@subfont\sf\rm}%
\def\psc{\@subfont\sc\rm}%
\def\ly{\fam\lyfam\twfvly}\textfont\lyfam\twfvly
   \scriptfont\lyfam\twfvly \scriptscriptfont\lyfam\twtyly
\@setstrut \rm}

\def\@xxxpt{}

\def\Huge{\@setsize\Huge{36pt}\xxxpt\@xxxpt}

\font\thtymi=cmmi10\@magscale6
    \skewchar\thtymi='177
\font\thtysy=cmsy10\@magscale6
    \skewchar\thtysy='60
\font\thtyly=lasy10\@magscale6
\font\thsirm=cmr12\@magscale6

\def\xxxvipt{\textfont\z@\thsirm
  \scriptfont\z@\thtyrm \scriptscriptfont\z@\twfvrm
\textfont\@ne\thtymi \scriptfont\@ne\thtymi \scriptscriptfont\@ne\twfvmi
\textfont\tw@\thtysy \scriptfont\tw@\thtysy \scriptscriptfont\tw@\twfvsy
\textfont\thr@@\tenex \scriptfont\thr@@\tenex \scriptscriptfont\thr@@\tenex
\def\unboldmath{\everymath{}\everydisplay{}\@nomath\unboldmath
        \textfont\@ne\thtymi \textfont\tw@\thtysy \textfont\lyfam\thtyly
        \@boldfalse}\@boldfalse
\def\boldmath{\@subfont\boldmath\unboldmath}%
\def\prm{\fam\z@\thsirm}%
\def\pit{\@subfont\it\rm}%
\def\psl{\@subfont\sl\rm}%
\def\pbf{\@getfont\pbf\bffam\@xxxpt{cmss12\@magscale6}}%
\def\ptt{\@subfont\tt\rm}%
\def\psf{\@subfont\sf\rm}%
\def\psc{\@subfont\sc\rm}%
\def\ly{\fam\lyfam\thtyly}\textfont\lyfam\thtyly
   \scriptfont\lyfam\thtyly \scriptscriptfont\lyfam\twfvly
\@setstrut \rm}

\def\@xxxvipt{}

\def\HUGE{\@setsize\HUGE{43pt}\xxxvipt\@xxxvipt}

\catcode`@=12

\font\tenex=cmex10 scaled 1200

\def\Sc#1{\hbox{\sc #1}}        
\font\oo=lcirclew10            

\def\bo{{\raise.05ex\hbox{\large$\Box$}\:}}             
\def\cbo{{\,\raise-.15ex\Sc [\,}}                       
\def\pa{\partial}                                       
\def\su{\sum}                                           
\def\TH{{\raise.2ex\hbox{$\displaystyle \bigodot$}\mskip-4.7mu \llap H \;}}
\def\face{\hbox{\normalsize$\;\;\:{\raise.9ex\hbox{\oo n}\mskip-13mu \llap
        {${\buildrel{\hbox{\frtnrm ..}}\over\smile}$}}\:$}}     
\def\Face{{\raise.2ex\hbox{$\displaystyle \bigodot$}\mskip-2.2mu \llap {$\ddot
        \smile$}}}                                      
\def\Lhat{{\bf\rlap{\kern-.09em$\hat{\phantom L}$}L}}
\def\Lcheck{{\bf\rlap{\kern-.09em$\check{\phantom L}$}L}}

\def\sp#1{{}^{#1}}                              
\def\sb#1{{}_{#1}}                              
\def\sl#1{\rlap{\hbox{$\mskip 1 mu /$}}#1}      
\def\leftrightarrowfill{$\mathsurround=0pt \mathord\leftarrow \mkern-6mu
        \cleaders\hbox{$\mkern-2mu \mathord- \mkern-2mu$}\hfill
        \mkern-6mu \mathord\rightarrow$}
\def\dvec#1{\vbox{\ialign{##\crcr
        \leftrightarrowfill\crcr\noalign{\kern-1pt\nointerlineskip}
        $\hfil\displaystyle{#1}\hfil$\crcr}}}           
\def\dt#1{{\buildrel {\hbox{\LARGE .}} \over {#1}}}     
\def\ddt#1{{\buildrel {\hbox{\LARGE .\kern-2pt.}} \over {#1}}}
\def\der#1{{\pa \over \pa {#1}}}                

\def\frac#1#2{{\textstyle{#1\over\vphantom2\smash{\raise.20ex
        \hbox{$\scriptstyle{#2}$}}}}}                   
\def\ha{\frac12}                                        
\def\sfrac#1#2{{\vphantom1\smash{\lower.5ex\hbox{\small$#1$}}\over
        \vphantom1\smash{\raise.4ex\hbox{\small$#2$}}}} 
\def\bfrac#1#2{{\vphantom1\smash{\lower.5ex\hbox{$#1$}}\over
        \vphantom1\smash{\raise.3ex\hbox{$#2$}}}}       
\def\afrac#1#2{{\vphantom1\smash{\lower.5ex\hbox{$#1$}}\over#2}}    

\def\boxes#1{
        \newcount\num
        \num=1
        \newdimen\downsy
        \downsy=-1.64ex
        \mskip-7.8mu
        \bo
        \loop
        \ifnum\num<#1
        \llap{\raise\num\downsy\hbox{$\bo$}}
        \advance\num by1
        \repeat}
\def\boxup#1#2{\newcount\numup
        \numup=#1
        \advance\numup by-1
        \newdimen\upsy
        \upsy=.82ex
        \mskip7.8mu
        \raise\numup\upsy\hbox{$#2$}}

\newskip\humongous \humongous=0pt plus 1000pt minus 1000pt
\def\caja{\mathsurround=0pt}

\newif\ifdtup
\def\panorama{\global\dtuptrue \openup2\jot \caja
        \everycr{\noalign{\ifdtup \global\dtupfalse
        \vskip-\lineskiplimit \vskip\normallineskiplimit
        \else \penalty\interdisplaylinepenalty \fi}}}
\def\li#1{\panorama \tabskip=\humongous                         
        \halign to\displaywidth{\hfil$\displaystyle{##}$
        \tabskip=0pt&$\displaystyle{{}##}$\hfil
        \tabskip=\humongous&\llap{$##$}\tabskip=0pt
        \crcr#1\crcr}}

\def\NP{Nucl. Phys. B}
\def\PL{Phys. Lett. }

\def\ref#1{$\sp{#1]}$}

\parskip=\medskipamount                 
\def\baselinestretch{1.2}       
\thicklines                         
\def\title#1#2#3#4{
\begin{document}
        {\hbox to\hsize{#4 \hfill QMW/PH/ #3}}\par
        \begin{center}\vskip.5in minus.1in {\Large\bf #1}\\[.5in minus.2in]{#2}
        \vskip1.4in minus1.2in {\bf ABSTRACT}\\[.1in]\end{center}
        \begin{quotation}\par}
\def\author#1#2{#1\\[.1in]{\it #2}\\[.1in]}
\def\AM{Aleksandar Mikovi\'c\,\footnote
   {Work supported by the U.K. Science and Engineering Research Council}
\footnote{E-mail address: MIKOVIC@V1.PH.QMW.AC.UK}
\\[.1in] {\it Department of Physics, Queen Mary and Westfield
College,\\ Mile End Road, London E1 4NS, U.K.}\\[.1in]}
\def\endtitle{\par\end{quotation}\vskip3.5in minus2.3in\newpage}

\catcode`@=12

\def\sect#1{\bigskip\medskip\goodbreak\noindent{\large\bf{#1}}\par\nobreak
        \medskip\markright{#1}}
\def\chsc#1#2{\phantom m\vskip.5in\noindent{\LARGE\bf{#1}}\par\vskip.75in
        \noindent{\large\bf{#2}}\par\medskip\markboth{#1}{#2}}
\def\Chsc#1#2#3#4{\phantom m\vskip.5in\noindent\halign{\LARGE\bf##&
        \LARGE\bf##\hfil\cr{#1}&{#2}\cr\noalign{\vskip8pt}&{#3}\cr}\par\vskip
        .75in\noindent{\large\bf{#4}}\par\medskip\markboth{{#1}{#2}{#3}}{#4}}
\def\chap#1{\phantom m\vskip.5in\noindent{\LARGE\bf{#1}}\par\vskip.75in
        \markboth{#1}{#1}}
\def\refs{\bigskip\medskip\goodbreak\noindent{\large\bf{REFERENCES}}\par
        \nobreak\bigskip\markboth{REFERENCES}{REFERENCES}
        \frenchspacing \parskip=0pt \renewcommand{\baselinestretch}{1}\small}
\def\unrefs{\normalsize \nonfrenchspacing \parskip=medskipamount}
\def\Item{\par\hang\textindent}
\def\Itemitem{\par\indent \hangindent2\parindent \textindent}
\def\makelabel#1{\hfil #1}
\def\topic{\par\noindent \hangafter1 \hangindent20pt}
\def\Topic{\par\noindent \hangafter1 \hangindent60pt}


\title{Hamiltonian Construction of W-gravity Actions}
{\AM}{91/13}{June 1991}
We show that all W-gravity actions can be easilly constructed and understood
from the point of view of the Hamiltonian formalism for the constrained
systems. This formalism also gives a method of constructing gauge invariant
actions for arbitrary conformally extended algebras.
\endtitle

Actions for a large class of W-gravity theories have been constructed so far
[1-6]. The method which was used for the construction of the gauge invariant
actions in all these cases was the Noether method and it's generalization in
terms of the $F\sb \pm$ auxiliary fields. It was pointed out in [6] that the
$F$ fields satisfy a formal bracket algebra reminiscent of a Poisson bracket
algebra in a Hamiltonian formalism where both $x\sp + = {1\over \sqrt 2}(x^0
+ x^1)$ and $x\sp - = {1\over \sqrt 2}(x^0 - x^1)$ are
regarded as evolution parameters. In the chiral case one can take $x^-$
(or $x^+$, which depends on the chirality of the currents) to be the evolution
parameter, and the usual Hamiltonian interpretation emerges [6].
A natural question then arises in the non-chiral case. What is the
first order Hamiltonian form of a W-gravity action? In this letter we give the
answer to this question,
and furthermore, due to the simplicity of the answer, we propose
an alternative way to that of reference [6]
of constructing gauge invariant actions for arbitrary
conformally extended algebras.

First we review those aspects of the Hamiltonian formalism which
will be needed for our construction.
Let $(p\sb i (t) , q\sp i (t) )$ be primary canonical variables of a
dynamical system with the Hamiltonian $H\sb 0 (p,q)$. Let
$G\sb \a (p,q)$ and $\Q\sb \m (p,q)$ be the first and the second
class constraints, respectively.
$t$ is the time, and the indices $i,\a,\m$ can take
both the discrete and the continious values, and can be bosonic or fermionic.
The Poisson bracket is defined as
$$ \{A,B \} = {\pa\sb L A\over\pa p\sb i}{\pa\sb R B\over\pa q\sp i}
- (-1)^{\e(A)\e(B)} {\pa\sb L B\over\pa p\sb i}{\pa\sb R A\over\pa q\sp i}
\eqno(1)$$
where $\pa\sb L$ and $\pa\sb R$ are left and right derivatives, while
$\e (X)=0,1$ if $X$ is a boson, fermion, respectively. Presence of the
second class constraints requires the Dirac bracket
$$ \{A,B \}\sb D = \{A,B \} - \{ A,\Q\sb \m \}(\D^{-1})\sp{\m\n}
\{\Q\sb \n , B \}\quad, \eqno(2)$$
where $\D\sb{\m\n}=\{\Q\sb \m ,\Q\sb \n \}$. The first class constraints
generate the gauge symmetries of the dynamical system and satisfy
$$ \{G\sb \a, G\sb \b \}\sb D = f\sb{\a\b}\sp \g G\sb \g \quad,\eqno(3)$$
$$ \{G\sb \a, H\sb 0 \}\sb D = h\sb{\a}\sp \b G\sb \b \quad,\eqno(4)$$
where equality is meant on the $\Q =0$ surface. $f$ and $h$ are
functions of the canonical variables.  The action is given by
$$ S = \int dt \left( p\sb i \dt{q}\sp i - H\sb 0 - \l\sp \a G\sb \a
\right)_{\Q =0}
\quad,\eqno(5)$$
where $\l\sp \a (t)$ are the Lagrange multipliers.
$S$ is invariant under the gauge transformations
$$ \li{ \d p\sb i = &\e\sp \a \{ G\sb \a , p\sb i \}\sb D \cr
        \d q\sp i = &\e\sp \a \{ G\sb \a , q\sp i \}\sb D \cr
        \d \l\sp \a = &\dt{\e}\sp \a - \l\sp \b \e\sp \g f\sb{\g\b}\sp \a
                      - \e\sp \b h\sb \b\sp \a \quad.&(6)\cr}$$
It is clear from (6) that the $\l\sp \a$ play the role of the gauge fields
corresponding to the symmetries generated by the $G\sb \a$.

Formula (5) together with (6) is exactly what one needs in order to construct
a gauge invariant action based on a given algebra. The only non-trivial
steps are finding the realization of the algebra in terms of the
canonical variables, and obtaining $H\sb 0$. This method was previously
used by Siegel [7], to construct gauge invariant actions for the
superparticle and the superstring. In the case of the $W$-algebras, obtaining
the canonical representation turns out to be easy, although there is a
subtlety in the choice of $H\sb 0$ in the
chiral case, related to the choice of the evolution
parameter.

Since one would like to a have a scalar field theory action
invariant under the $W$
gauge transformations, the canonical coordinates are then the two dimensional
scalar fields
$\f\sb i (\s,\t)$, where $i=1,...,n$ and $\t$ is the time ($\t = x^0$,
$\s = x^1$). Let
$P\sb i (\s,\t)$ be the canonically conjugate momenta, satisfying
$$ \{ P\sb i (\s,\t), \f\sb j (\s^{\prime},\t)\} =
\d\sb{ij}\d (\s - \s^{\prime}) \quad.
\eqno(7)$$
In order to construct the action, we need a canonical representation of
a given W-algebra.
This can be obtained from the free field representation of
a $W$-algebra currents [4], by replacing $\pa\sb{\pm}\f\sb i$ with
$$ \hat{P}\sb i\sp{\pm} = {1\over\sqrt 2}(P\sb i \pm \f^{\prime}\sb i ) \quad,
\eqno(8)$$
where $f^{\prime} = \der{\s}f$, and $f^{(n)}=(\der{\s})^n f$.
Then the Poisson bracket alegbra of the $\hat{P}$'s is isomorfic to
that of the $\pa\f$'s
$$ \{ \hat{P}\sb i\sp \pm (\s,\t), \hat{P}\sb j\sp \pm (\s^{\prime},\t)\} =
\mp\d\sb{ij}\d^{\prime} (\s - \s^{\prime}) \quad. \eqno(9)$$
Hence the generators of the $W\sb N$ algebra are given as
$$ W\sb{\pm s} = \frac1s d\sb{i\sb 1 ...i\sb s}\hat{P}\sb{i\sb 1}\sp \pm
\cdots\hat{P}\sb{i\sb s}\sp \pm \quad (s=2,...,N)\quad. \eqno(10)$$
The Poisson bracket algebra of the constraints (10) closes if the constants
$d\sb{i\sb 1 ...i\sb s}$
satisfy certain algebraical relations [4],[8].
In the $W\sb 3$ case these are
$$ d\sb{(ij|k}d\sp k\sb{|l)m} = \k d\sb{(ij}d\sb{l)m}\quad,\eqno(11)$$
and the Poisson bracket algebra is
$$\li{\{T\sb \pm (\s),T\sb \pm (\s^{\prime})\}
&= \mp\d^{\prime} (\s - \s^{\prime}) (T\sb \pm (\s) +
T\sb \pm (\s^{\prime})) \cr
\{T\sb \pm (\s),W\sb \pm (\s^{\prime})\}
&= \mp\d^{\prime} (\s - \s^{\prime}) (W\sb \pm (\s) +
2W\sb \pm (\s^{\prime})) \cr
\{W\sb \pm (\s),W\sb \pm (\s^{\prime})\}
&= \mp 2\k\d^{\prime} (\s - \s^{\prime}) (T\sp 2\sb \pm (\s) +
T^2\sb \pm (\s^{\prime})) \quad,&(12)\cr}$$
where $T=W\sb 2$ is the energy-momentum tensor and $W=W\sb 3$.

In the non-chiral case the theory is diffeomorphism invariant, and therefore
$H\sb 0 =0$ (otherwise, the wave functional $\Psi[\f]$ would depend explicitely
on the unphysical
parameter $\t$, since $i\der{\t}\Psi = \hat{H}\sb 0 \Psi $). Then according to
(5) the gauge invariant action is simply
$$ S\sb N = \int d\s d\t \left( P\sb i \dt{\f}\sb i - h\sp{\pm} T\sb{\pm}
- \su_{s=3}^{N} B\sp{\pm}\sb s W\sb{\pm s}\right) \quad,\eqno(13)$$
where $h$ and $B$ are the
lagrange multipliers, which are also the gauge fields corresponding to
the $W$-symmetries. The gauge transformation laws can be determined from
(6). In the $W\sb 3 $ case they become
$$\li{\d P\sb i =& {1\over\sqrt2} ( \e\sp + \hat{P}\sb{+i}
- \e\sp - \hat{P}\sb{-i})^{\prime}
 + {1\over\sqrt2}( \x\sp + d\sb{ijk}\hat{P}\sb +\sp j \hat{P}\sb{+}\sp k -
\x\sp - d\sb{ijk}\hat{P}\sb -\sp j \hat{P}\sb{-}\sp k
)^{\prime} ,&(14.a)\cr
\d \f\sb i =& {1\over\sqrt2} \e\sp \pm \hat{P}\sb{\pm i}  +
{\x\sp \pm \over\sqrt2}d\sb{ijk}
\hat{P}\sb{\pm}\sp{ j} \hat{P}\sb{\pm}\sp{ k}
\quad,&(14.b)\cr
\d h\sp \pm =& \dt{\e}\sp \pm \mp h\sp \pm (\e\sp \pm)^{\prime} \pm
(h\sp \pm )^{\prime}\e\sp \pm \pm 2\k (\x\sp \pm (B\sp \pm)^{\prime} -
(\x\sp \pm)^{\prime} B\sp \pm ) T\sb \pm \quad,&(14.c)\cr
\d B\sp \pm =& \dt{\x}\sp \pm \pm 2(h\sp \pm)^{\prime}\x\sp \pm
\mp h\sp \pm (\x\sp \pm )^{\prime} \mp 2 B\sp \pm (\e\sp \pm)^{\prime}
\pm (B\sp \pm )^{\prime}\e\sp \pm \quad,&(14.d)\cr}$$
where we have taken $d\sb{ij}=\d\sb{ij}$, and $\e^\pm$ are the parameters of
the $T\sb \pm$ transformations, while $\x^\pm$ are the parameters of the
$W\sb \pm$ transformations.

In order to find a geometrical interpretation of the action (13) we need
to know it's second order form.
It can be obtained by substituting in the expressions for the
momenta $P\sb i$ obtained from the equation of motion
$${\d S\over \d P\sb i} = 0\quad.\eqno(15)$$
In the $W\sb 3$ case one gets
$$ \dt{\f}\sb i - \ha h\sp \pm (P\sb i \pm \f^{\prime}\sb i ) -
{B\sp \pm \over 2\sqrt2}d\sb{ijk}
(P\sb j \pm \f^{\prime}\sb j )(P\sb k \pm \f^{\prime}\sb k ) = 0
\quad.\eqno(16)$$
This is a quadratic equation in the $P$'s, and therefore the second order
form of the Lagrangian density
will be a non-polynomial function of $\pa\f$, $h$ and $B$, which can be written
as an infinite power series in those variables.
This will be a generic feature of all
W-gravity actions, except for the $W\sb 2$ case (usual 2d gravity), where the
equation (15) is linear in $P$'s ($B$ indipendent part of (16)).
This is analogous to the results of the Noether procedure,
where a pair of
auxilliary fields is introduced in order to have a closed form
of the action, and they satisfy quadratic equations.
However, the advantage of the Hamiltonian formalism is that one is
automatically provided with the closed form of the action from the begining,
and the
momenta are the auxilliary fields. The standard drawback is the loss of
manifest covariance.

As we have explained, the second order form of the action can be explicitely
evaluated in the $W\sb 2$ case
$$S\sb 2 =\ha\int d^2 \s \sqrt{-g}g\sp{\m\n} \pa\sb \m\f\sp i \pa\sb \n\f\sb i
\quad,\eqno(17)$$
where
$$ \tilde{g}\sp{00} = {2\over h\sp + + h\sp -} \quad,\quad
\tilde{g}\sp{01} = {h\sp - - h\sp + \over h\sp + + h\sp -} \quad,\quad
\tilde{g}\sp{11} = - {2h\sp + h\sp -\over h\sp + + h\sp -} \quad,\eqno(18)$$
and $\tilde{g}\sp{\m\n} = \sqrt{-g}g\sp{\m\n}$. The geometrical
interpretation of (17) is that it represents the action for 2d gravity coupled
to scalars. This can be verified by noticing first that the transformation law
for a field $\f\sb i$ ((14.b) with $\x =0$) can be written as
$$ \d\f\sb i = \e\sp \pm\tilde{e}\sb \pm\sp \m \pa\sb \m \f\sb i =
\e\sp \m \pa\sb \m \f\sb i \quad,\eqno(19)$$
where
$$ \tilde{e}\sb{\pm}\sp \m = {1\over h\sp + + h\sp -}
\pmatrix{1 &h\sp - \cr 1 &-h\sp +\cr} \quad.\eqno(20)$$
Equation (19) is an infinitesimal
diffeomorphism transformation of a scalar, where $\e\sp \m$ is the parameter
of the transformation. Then one can show that the components of
$\tilde{g}\sp{\m\n}$ given by
(18) transform under the $\x =0$ part of (14.c) as
$$\d\tilde{g}\sp{\m\n} = \pa\sb \r (\e\sp \r \tilde{g}\sp{\m\n}) -
\pa\sb \r \e\sp{(\m|}\tilde{g}\sp{|\n)\r} \quad,\eqno(21)$$
which is the
diffeomorphism transformation of a densitized metric
generated by the parameter $\e\sp \m$. The metric $g\sp{\m\n}$ can be written
as
$$g\sp{\m\n}= {\r\over (h\sp + + h\sp - )^2}
\pmatrix{2 &h\sp - - h\sp +\cr h\sp - - h\sp + &-2h\sp + h\sp - \cr}
= e\sb +\sp{(\m|} e\sb -\sp{|\n)} \quad, \eqno(22)$$
where $e\sb \pm\sp \m =\sqrt{\r}\tilde{e}\sb \pm\sp \m$ are the zweibeins
and $\r$ is the conformal mode of the
metric. Note that the action (17) is indipendent of $\r$ due to the Weyl
symmetry
$$ \d g\sp{\m\n} = \o g\sp{\m\n} \eqno(23) $$
so that
$$ \sqrt{-g} = |e|= {h\sp + + h\sp - \over\r} \quad. \eqno(24) $$

In the $W\sb 3$ case the geometric interpretation is not obvious
due to the lack of the explicit form of the second order action.
In order to obtain the second order form of the action we will rewrite
the momentum equation (16) as
$$\li{ P\sb i &= {2\over h\sp + + h\sp -}\left[ \dt{\f}\sb i -
\ha (h\sp + - h\sp - )\f^{\prime}\sb i -
{B\sp \pm \over\sqrt2}d\sb{ijk}
\hat{P}\sb \pm\sp j \hat{P}\sb \pm\sp k \right] \cr
&= P\sb i\sp 0 + P\sb i\sp 1 \quad, &(25)\cr}$$
where $P\sb i\sp 0$ is the $B$ indipendent part of (25). Then the action (13)
takes the following form
$$ S\sb 3 = \int d^2 \s \left( |e| \pa\sb + \f\sb i \pa\sb -\f\sb i
-{h\sp + + h\sp - \over 4}(P\sb i\sp 1 )^2
-{ B\sp{\pm}\over 3}d\sb{ijk}\hat{P}\sb{\pm}\sp i
\hat{P}\sb{\pm}\sp j \hat{P}\sb{\pm}\sp k \right) \quad,
\eqno(26)$$
where $\pa\sb{\pm}= e\sb{\pm}\sp \m \pa\sb \m$. $\hat{P}$ satisfies
$$ \hat{P}\sb{\pm i} = \sqrt{2} \tilde{\pa}\sb \pm\f\sb i -
{B\sp \pm \over h\sp + + h\sp -}d\sb{ijk}
\hat{P}\sb \pm \sp j \hat{P}\sb \pm \sp k \quad,\eqno(27)$$
where $\tilde{\pa} = \r^{-\ha}\pa$.
By using (27) one can obtain the power series expansion of $\hat{P}$ in terms
of $\pa\sb \pm\f$ and $B$, which can be inserted into (26) to give the
corresponding power series expansion of the action. Up to the first order in
$B$ the Lagrange desity can be written as
$$ \cl = |e| \left( \pa\sb + \f\sb i \pa\sb -\f\sb i
- B\sp{\pm\pm\pm} d\sb{ijk}\pa\sb{\pm}\f\sp i
\pa\sb{\pm}\f\sp j \pa\sb \pm\f\sp k \right) + O(B^2)\quad,
\eqno(28)$$
where
$$ B\sp{\pm\pm\pm} = {2\sqrt2\over 3}{B\sp \pm \over \sqrt{\r}(h\sp +
+ h\sp - )}\quad.\eqno(29)$$
Equation (28) suggests the geometrical interpretation based on the
zweibeins introduced in the $W\sb 2$ case. However, that is not possible,
since the second order in $B$ contribution to $\cl$ is
$$ \cl\sb{(2)} ={9\over4} |e| \left(
B\sp{+++} d\sb{ijk}\pa\sb{+}\f\sp j \pa\sb +\f\sp k +
B\sp{---} d\sb{ijk}\pa\sb{-}\f\sp j \pa\sb -\f\sp k \right)^2 \quad,
\eqno(30)$$
which is not diffeomorphism invariant. The reason for this is that the $\pm$
indicies in (28) and (30) are not covariant because the diffeomorphism
transformation of $\f\sb i$ in the $W\sb 3$ case is not $\d\f\sb i =\e\sp \pm
\tilde{\pa}\sb \pm\f\sb i$ but
$$\d\f\sb i ={1\over\sqrt2}\e\sp \pm\hat{P}\sb{\pm i} =
\e\sp \pm \left( \tilde{\pa}\sb \pm\f\sb i -
{\sqrt2 B\sp \pm \over (h\sp + + h\sp -)}d\sb{ijk}\tilde{\pa}\sb \pm\f\sp j
\tilde{\pa}\sb \pm\f\sp k + \cdots \right) \quad.\eqno(31)$$
Note that the equation (31) can be rewritten as
$$\d\f\sb i =\e\sp \m \pa\sb \m\f\sb i \quad,
\quad \e\sp \m = f\sp \m (\e\sp \pm ,h\sp \pm ,B\sp \pm ,\pa\sb \pm\f )
\quad.\eqno(32)$$
The function $f$ is not unique, and
the difficulty at the moment is to see which choice of
$f$ is the right one. However, given that the form (32) exists, and
that $\cl$ can be written as
$$\li{ \cl =& \ha\tilde{g}\sp{\m\n}\pa\sb \m \f\sp i \pa\sb \n \f\sb i
+  \tilde{B}\sp{\m\n\r}d\sb{ijk}
\pa\sb \m \f\sp i \pa\sb \n \f\sp j \pa\sb \r\f\sp k \cr
&+ \tilde{C}\sp{\m\n\r\h}d\sb{ij}\sp{m}d\sb{mkl}
\pa\sb \m \f\sp i \pa\sb \n \f\sp j \pa\sb \r\f\sp k \pa\sb \h \f\sp l
+ \cdots \quad,&(33)\cr}$$
one can argue that the objects $\tilde{g}$, $\tilde{B}$, $\tilde{C}$,
... , must transform
as tensor densities in order for (33) to be invariant under (32).
As far as the zweibein interpretation is concerned, the formula (32) gives a
clue. Namely, one can write
$$\e\sp \m = f\sp \m (\e\sp \pm ,h\sp \pm ,B\sp \pm ,\pa\sb \pm\f ) =
\e\sp \pm \tilde{e}\sb \pm\sp \m (h\sp \pm ,B\sp \pm ,\pa\sb \pm\f )
\quad,\eqno(34)$$
which means that the zweibeins (20) change in the $W\sb 3$ case.
They become functions of $B$ and $\pa\f$ such that when $B\to 0$ then
$\tilde{e}(h,B,\pa\f) \to \tilde{e}(h)$ of (20).

Besides the diffeomorphism invariance, the generalized Weyl symmetry [4],[9]
is also
obscured. Heuristically, it is there by construction, since we used only four
indipendent gauge fields $h\sp \pm$ and $B\sp \pm$. The fields $\tilde{g}$,
$\tilde{B}$, $\tilde{C}$, ... in (33) are functions of $h$ and $B$, and one
can check order by order in $\pa\f$ that
$$ \tilde{g}\sb{\m\n}\tilde{B}\sp{\m\n\r} = 0 \quad,\quad
   \tilde{C}\sp{\m\n\r\s} = \tilde{g}\sb{\c\h}\tilde{B}\sp{\c\m\n}
\tilde{B}\sp{\h\r\s} \quad, \eqno(35)$$
and so on.
The generalized Weil invariance of the complete action follows from the results
obtained recently by Hull in the context of the $W\sb{\infty}$ gauge theory [9].
Namely, a Lagrange density $\cl (x,\f(x),\pa\f(x))$ is invariant under the
generalized Weyl symmetry if the function
$$ \tilde{F}(x,y) = \cl (x,\f(x),\pa\f(x))|_{\pa\f =y} \eqno(36) $$
satisfies the Monge-Ampere equation
$$ det\left|{\pa^2 \tilde{F}(x,y)\over \pa y\sb \m \pa y\sb \n}\right| = -1
\quad.\eqno(37)$$
One can show that
$$\tilde{F}(x,y) = P\dt{\f} - f(x,P + \f^{\prime}) -\bar{f}(x,P-\f^{\prime})
\eqno(38)$$
is a general solution of (37), where $P$ is the auxilliary field determined
by $P={\pa \tilde{F}\over \pa\dt{\f}}$,
and $f$ and $\bar{f}$ are arbitrary functions.
Note that the expression (38) is exactly the hamiltonian form of the Lagrange
density for a $W$-gravity theory (13), and the functions $f$ and $\bar{f}$
are just the linear combinations of the constraints (10).

It is clear that given any extended conformal algebra realized in terms of
free scalar fields, the gauge invariant action can be written down imidiately
by following the procedure we have presented. Even when the classical Poisson
bracket algebra of the constraints
contains a central extension, the method still works, since
then the central charge can be treted as an extra Abelian generator, and the
formula (5) would still apply. In that case the free field expressions for the
currents will contain the higher derivative terms $\pa^n \f$ [8], which would be
replaced by $\hat{P}^{(n-1)}$ in order to obtain the constraints.
Note that this construction is very similar to the corresponding
Noether construction in terms of the auxilliary $F$-fields [6], where
$\pa^n \f$ is replaced with $\pa^{n-1}F$. However, there is a substantial
difference. While the Hamiltonian action will still be first order in time
derivatives, the Noether action will contain higher order time derivatives.
This might be an indication that in the case when the central charges
(or equivalently the background charges) are present,
the two methods give
different actions. This requires a further investigation.

Inclussion of the fermions in the Hamiltonian formalism is straightforward.
They are represented as the Grassman valued fields $\j\sp \pm\sb a (\s,\t)$,
which are canonically self-conjugate
$$\{\j\sp \pm\sb a (\s,\t),\j\sp \pm\sb b (\s^{\prime},\t)\}\sb D =
\d\sb{ab}\d (\s -\s^{\prime}) \quad.\eqno(39)$$
The Hamiltonian constraints associated with a given extended conformal algebra
can be obtained from a free-field representation by replacing
$\pa\sb \pm\sp n\f$ with $\hat{P}\sp{(n-1)}\sb \pm$ and
$\pa\sb \pm\sp n\j\sb \pm$ with $\j\sp{(n)}\sb \pm$, so that the
corresponding gauge invariant action is
$$ S\sb G = \int d\s d\t \left[ P\sb i \dt{\f}\sb i +
\ha\j\sp \pm\sb a\dt{\j}\sp \pm\sb a - \cb\sp{\pm}\sb \a G\sb{\pm \a}
(\hat{P}\sb \pm,\hat{P}^{\prime}\sb \pm,...,\j\sb \pm ,\j^{\prime}\sb \pm ,...
)\right]\quad,\eqno(40)$$
where $G\sb{\pm \a}$ are the left/right currents, satisfying (3), and
$\cb\sp \pm\sb \a$ are the corresponding gauge fields (Lagrange multipliers).
The gauge transformations can be determined from (6).

Our final comment concernes the chiral case. As shown in [6], the Noether
form of the chiral action [1],[4], follows directly from the formula (5) if the
evolution parameter is chosen to be $x^-$. In the case when $x^0$ is chosen
as the evolution parameter, the form of the chiral action changes. By analysing
the $W\sb 2$ case one can see that a non-zero $H\sb 0$ appears
$$ H\sb 0 = -\int d\s P\sb i \f^{\prime}\sb i \quad,\eqno(41)$$
which is a consequence of the fact that the chiral theory is not diffeomorphism
invariant (i.e. only a ``half" of the diffeomorphism invariance is present).
Since
$$ \{ W\sb{+s} (\s) , H\sb 0 \} = \der{\s} W\sb{+s} (\s) \quad,\eqno(42)$$
so that (4) is satisfied, the action is given by (5)
$$ S_{N}^{(ch)} = \int d\s d\t \left( P\sb i \dt{\f}\sb i + P\sb i \f^{\prime}
\sb i - h\sp{+} T\sb{+}
- \su_{s=3}^{N} B\sp{+}\sb s W\sb{+ s}\right) \quad. \eqno(43)$$
The gauge transformations can be read off from (6). For $N\ge 3$
the second order form of this action appears to be nonpolynomial, in contrast
to the polynomial chiral Noether action. However, one can expect from
the general arguments that the Hamiltonian form of the chiral
Noether action is precisely (43),
which implies that the second order form of (43) is polynomial.

I would like to thank Chris Hull for helpfull discussions.

\refs
\Item{[1]}C.M. Hull, \PL 240B (1990) 110
\Item{[2]}K. Schoutens, A. Sevrin and P. van Nieuwenhuizen, \PL 243B (1990) 245
\Item{[3]}E. Bergshoeff, C.N. Pope, L.J. Romans, E. Sezgin, X. Shen and K.S.
          Stelle, \PL 243B (1990) 350
\Item{[4]}C.M. Hull, \NP 353 (1991) 707
\Item{[5]}C.M. Hull, \PL 259B (1991) 68
\Item{[6]}A. Mikovi\'c, \PL 260B (1991) 75
\Item{[7]}W. Siegel, \NP 263 (1985) 93
\Item{[8]}L.J. Romans, \NP 352 (1991) 829
\Item{[9]}C.M. Hull, The geometry of W-gravity, QMWC preprint (1991)
          QMW/PH/91/6

\end{document}

